\documentclass[review]{elsarticle}
\usepackage{graphicx}
\usepackage{dcolumn}
\usepackage{bm}
\usepackage{longtable}
\usepackage{amsmath}
\usepackage{amsfonts,amssymb}
\usepackage{mathtools,amsthm,adjustbox}
\usepackage{lipsum}
\usepackage[margin=2.9cm]{geometry}
\def\a{\alpha} \def\e{\hat\eta}

\def\dsy{\displaystyle}
\def\de{\delta}

\def\bn{\begin{equation}}
\def\en{\end{equation}}
\def\bny{\begin{eqnarray}}
\def\eny{\end{eqnarray}}
\def\be{\begin{eqnarray*}}
\def\ee{\end{eqnarray*}}
\def\bc{\begin{center}}
\def\ec{\end{center}}
 
\def\p{\partial}
\def\({\left(}
\def\){\right  )}
\def\[{\left[}
\def\]{\right]}
\def\bc{\begin{center}}
\def\ec{\end{center}}

\def\ds{\displaystyle}

\newtheorem{dfn}{Definition}[section]
\newtheorem{thm}{Theorem}[section]
\newtheorem{rem}{Remark}[section]
\newtheorem{pro}{Proposition}[section]

\newtheorem{cor}{Corollary}[section]
\newtheorem{lem}{Lemma}[section]
\newtheorem{exm}{Example}[section]

\def\bn{\begin{equation}}
\def\en{\end{equation}}
\def\bny{\begin{eqnarray}}
\def\eny{\end{eqnarray}}
\def\be{\begin{eqnarray*}}
\def\ee{\end{eqnarray*}}
\def\bdn{\begin{dfn}}
\def\edn{\end{dfn}}
\def\btm{\begin{thm}}
\def\etm{\end{thm}}
\def\bpf{\begin{proof}}
\def\epf{\end{proof}}
\def\bpn{\begin{pro}}
\def\epn{\end{pro}}
\def\brk{\begin{rem}}
\def\erk{\end{rem}}
\def\bcy{\begin{cor}}
\def\ecy{\end{cor}}
\def\blm{\begin{lem}}
\def\elm{\end{lem}}
\def\bex{\begin{exm}}
\def\eex{\end{exm}}
\def\e{\epsilon}

\usepackage{lineno,hyperref}
\modulolinenumbers[5]

\journal{Journal of Geometry and Physics}

\usepackage{numcompress}

\begin{document}

\begin{frontmatter}

\title{Group invariant transformations for the Klein-Gordon equation in three dimensional flat spaces}

\author[]{Sameerah Jamal \corref{mycorrespondingauthor}}
\address{School of Mathematics and Centre for Differential Equations,Continuum Mechanics and Applications,
University of the Witwatersrand, Johannesburg, South Africa}
\ead{Sameerah.Jamal@wits.ac.za}
\cortext[mycorrespondingauthor]{Corresponding author}

\author{Andronikos Paliathanasis}
 \address{Instituto de Ciencias F\'{\i}sicas y Matem\'{a}ticas, Universidad Austral de
Chile, Valdivia, Chile}
\address{Institute of Systems Science, Durban University of Technology, PO Box 1334,
Durban 4000, Republic of South Africa}
\ead{ anpaliat@phys.uoa.gr}

\begin{abstract}
We perform the complete symmetry classification of the Klein-Gordon equation in maximal symmetric spacetimes.
The central idea is to find all possible potential functions $V(t,x,y)$ that admit Lie and Noether symmetries. This is done by using the relation between the symmetry vectors of the differential equations and the elements of the conformal algebra of the underlying geometry. For some of the potentials, we use the admitted Lie algebras to determine corresponding invariant solutions to the Klein-Gordon equation. An integral part of this analysis is  the problem of  the classification of Lie and Noether point symmetries of the wave equation.\end{abstract}


\end{frontmatter}


\section{\label{sec:level1}Introduction}
The group classification problem was initiated by  Ovsiannikov \cite{osv} who analysed the nonlinear heat equation. Since then numerous studies have been devoted to group classifications of fundamental equations that model mathematical, relativistic, biological and  physical phenomena \cite{ss0,ss1,ss2,ss3,ss4,ss5,j6} . In addition, there is now a rich body of literature surrounding Lie symmetry theory, its scheme and vast applications to differential equations \cite{o,b1,b2,b3,b4}.  In particular, wave and Klein-Gordon equations are of particular interest as they are two important equations in all areas of physics. A knowledge of the Lie symmetry structures of the Klein-Gordon equation in a Riemannian space enables the determination of solutions of this equation which is invariant under a given Lie symmetry. 

Indeed, recent investigations \cite{j1,j5,da,pal,j3,j4,j7} have revolved around wave, Klein-Gordon, Poisson and Schr\"{o}dinger equations - showing that the Lie symmetry vectors are obtained directly from the collineations of a metric which defines the underlying geometry in which the evolution occurs. In \cite{tp} it was proved that for a linear (in derivatives) second-order partial differential equation (PDE), the Lie point symmetries are related to the conformal algebra of the geometry defined by the PDE. In \cite{pal2}, a geometric approach related the Lie symmetries of the Klein-Gordon equation to the conformal algebra of classes of the Bianchi I spacetime and a study of potential functions  was performed. Whilst recently the connection between collineations and symmetries was established for a system of quasilinear PDEs  \cite{newAP}. A similar result has been proved for the Poisson equation  \cite{poisson}.

In this paper, we use geometric results that transfer the problem of the Lie and Noether symmetry classification of the Klein-Gordon equation to the problem of determining the conformal Killing vectors that  admit appropriate potential functions. Inspired from the symmetry classification of the two- and three-dimensional Newtonian systems in which a geometric approach was applied \cite{ss6,ss7}, in this work in order to perform the classification, the conformal Killing vectors of the space are used to solve a constraint condition. The general results are applied to two practical problems viz.,  the classification of all potential functions in a three dimensional Euclidian and Minkowski space, for which the  Klein-Gordon equation admits Lie and Noether point symmetries and secondly, the Lie point symmetries are used to determine invariant solutions of the equation.

The paper is organized as follows. Section 2 provides the geometrical preliminaries and the theoretical background about symmetry analysis. In section 3, we  state the main theorem containing the constraint condition. Section 4 provides a short review about the spacetime and its properties and we perform the symmetry classification for the
Klein-Gordon equation and the corresponding potentials.  Section 5 illustrates some invariant solutions for the Klein-Gordon equation using  particular potential functions. Finally in section 6 we draw our conclusions.

\section{Preliminaries}
In this section we review the definitions and properties of spacetime  collineations and of the point symmetries of differential equations.
\setcounter{equation}{0}
\subsection{Lie and Noether Point Symmetries}
Consider a  system with  $q$  unknown functions $u^a$ which depends on $p$ independent variables  $x^i$, i.e. we denote
$u = (u^1,\ldots, u^q)$ and $x= (x^1,\ldots, x^p)$, respectively.
Let \bn\label{main} G_\alpha\(x,u^{(k)}\)=0, \quad \alpha=1, \ldots, q,\en
be a system of $m$ nonlinear differential equations, where $u^{(k)}$ represents the $k^{th}$  derivative of $u$ with respect to $x$.
A one-parameter Lie group of transformations ($\varepsilon$ is the group parameter) that is invariant under (\ref{main}) is given by \bn\label{main2} \bar{x}=\Xi (x,u;\varepsilon)\quad \bar{u}=\Phi (x,u;\varepsilon).\en
Invariance of (\ref{main})  under the transformation (\ref{main2})  implies that any solution $u = \Theta(x)$ of  (\ref{main}) maps into another solution $v = \Psi(x; \varepsilon) $ of  (\ref{main}).
 Expanding (\ref{main2}) around the identity $\varepsilon = 0$, we can generate the following infinitesimal transformations:
\bn\begin{array}{ll}
&\bar{x}^i=x^i+\varepsilon\xi^i(x,u)+{\cal O}(\varepsilon^2),\quad i=1, \ldots, p,\\
&\bar{u}^\alpha=u^\alpha+\varepsilon\eta^\alpha(x,u)+{\cal O}(\varepsilon^2). 
\end{array}
\en
The action of the Lie group can be recovered from that of its
infinitesimal generators acting on the space of independent and dependent variables. Hence, we consider the following vector field
\bn\label{xop}\dsy{X=\xi^i\p_{x^i}+\eta^\alpha\p_{u^{\alpha.}}}\en
The action of X is extended to all derivatives appearing in the equation in question through the appropriate prolongation. The infinitesimal criterion for invariance is given by \bn\label{crib}X\[\textrm{LHS Eq.} (\ref{main}) \]\mid_{Eq. (\ref{main}) }=0.\en Eq.  (\ref{crib}) yields an overdetermined system of linear homogeneous equation which can be solved algorithmically, more details can be found in  \cite{o} among other texts.

The generalized total differentiation operator $D_i$ with respect to $x^i$ is given by
\begin{equation} \label{eq1}
D_i={{\p}\over{\p x^i}}+u_i^\a {{\p}\over{\p u^\a}}+
u_{ij}^\a {{\p}\over{\p u_j^\a }} + \ldots.
\end{equation}
and $W^\a$ is the characteristic function given by
\begin{equation}\label{eq8}
W^\a=\eta^\a-\xi ^ju_j^\a\;.\end{equation}
The Euler-Lagrange equations,
if they exist, is the system
${\de}L/{\de u^\a}=0$, where $\de/\de u^\a$ is the
Euler-Lagrange operator given by \bn {{\de}\over{\de
u^\a}}={{\p}\over{\p u^\a}}+\sum_{s\geq1}(-1)^sD_{i_1}\cdots
D_{i_s}\;{{\p}\over{\p u^\a_{i_1\cdots i_s}}}\;. 
\label{3.12} \en $L$ is referred to as a Lagrangian.
If we include point dependent gauge terms $f_1, \ldots, f_n$, the Noether symmetries
$X$ are given by
\bn\label{eq13n}X(L) + LD_i(\xi^i) = D_i(f_i).\en
 \subsection{Conservation Laws}
 Corresponding to each $X$, there exists a conserved vector $(T^1,\ldots,T^n)$ that may then be determined by
 Noether's theorem \cite{noet}
\begin{equation}\label{con-law-T}
T^i= f^i - N^i (L). 
\end{equation}
where
\begin{equation}\label{con-law}
D_i \,  T^i=0
\end{equation}
along the solutions of the differential equation.
 Here $N^i$ is the Noether operator associated with the symmetry operator $X$ given by
\begin{equation}\label{N-op}
{\bf N^i}=\xi ^i + W^\a {{\de}\over{\de
u^\a}}+ \sum_{s\ge1}D_{i_1}\ldots D_{i_s}{W}^\a \;{{\de}\over{\de
u_{i i_1\ldots i_s}^\a}},\end{equation}
where $\de/\de u^\a$ is the Euler-Lagrange operator given by (\ref{3.12}).
\subsection{Invariant Solutions}
The operator in Eq. (\ref{xop}) can be used to define the Lagrange system
$$\frac{dx^i}{\xi^i}=\frac{du}{\eta}=\ldots$$
whose solution provides the invariant functions
\bn\label{in}W^{[r]}(x^i,u).\en
These invariants can be used in order to reduce the order of the PDE. Further details of the relevant equations and formulae can be found in, inter alia, \cite{hs}.
\subsection{Collineations of Riemannian Spaces}
A one-parameter group of conformal motions, or Conformal Killing vectors (CKVs), generated by the vector field  $X$ is defined by \cite{kat}
\bn\label{ck} {\cal L}_Xg_{ab}=2\psi g_{ab},\en where ${\cal L}_X$ is the Lie derivative operator along the vector field $X$ and $\psi=\psi(x^a)$ is a conformal factor. If $\psi_{;ab}\ne0$, the CKV is said to be proper.  The possible cases of $\psi$ provide special cases which form subalgebras on the conformal algebra of the space:
\bn\begin{array}{lcl}\label{ckv}
\psi_{;ab}=0&\Longleftrightarrow&X\, \textrm{is a special CKV (sCKV)},\\
\psi_{,a}=0&\Longleftrightarrow&X\, \textrm{is a homothetic vector (HV)},\\
\psi=0&\Longleftrightarrow&X\, \textrm{is a Killing vector (KV)}.
\end{array}\en

\section{Klein-Gordon Equations}
The linear second order in derivatives  Klein-Gordon equation is expressed as
\bn\begin{array}{ll}\square u+V(x^i)u=&{1 \over {\sqrt {
|-g|}}}{\p \over \p x^i} \left({\sqrt{|-g|}}g^{ik}{\p u \over \p x^k}
\right)\\&+V(x^i)u=0, \label{kleinGordon}\end{array}\en
where $\square$ is the d'Alembertian operator.
In a recent paper by Paliathanasis and Tsamparlis \cite{pal}, it was shown that the Lie point symmetries of the Klein-Gordon equation in a general Riemannian space are elements of the conformal algebra of the space, modulo a constraint relation involving the Lie symmetry vector and the potential entering the Klein-Gordon equation. More specifically the following theorem was proved.

\textbf{Theorem 1.} The Lie point symmetries of the Klein-Gordon equation (\ref{kleinGordon}) in a Riemannian space of dimension $n$ are generated from the elements of the conformal algebra of the metric, as follows:
\begin{enumerate}
\item For $n>2$ the Lie symmetry vector is
 $$X=\xi^i(x^k)\p_i+\(\frac{2-n}{n}\psi(x^k)u+a_0 u+b(x^k)\)\p_u$$
 where $\xi^k$ is a CKV with conformal factor $\psi(x^k), b(x^k)$  is a solution of (\ref{kleinGordon}) and the following condition involving the potential
 \bn\label{pa} \xi^kV_{,k} + 2\psi V-\frac{ 2-n}{2}\triangle\psi = 0. \en
 \item For $n=2$ the Lie symmetry vector is
 $$X=\xi^i(x^k)\p_i+\(a_0 u+b(x^k)\)\p_u$$
 where $\xi^k$ is a CKV with conformal factor $\psi(x^k), b(x^k)$  is a solution of (\ref{kleinGordon}) and the following condition involving the potential
 \bn\label{p} \xi^kV_{,k} + 2\psi V = 0. \en
\end{enumerate}
Constraint condition (\ref{pa}) acts as a double selection rule - selecting for each CKV a corresponding potential or, if this is not possible, abandoning the CKV for being a Lie point symmetry of the Klein-Gordon equation \cite{pal2}.
We remark that the case $V(x^i)=0$ reduces the Klein-Gordon equation to that of the wave equation. Further, it is well known that a Lie algebra contains the Noether algebra, where the Noether algebra will exclude a  symmetry involving the dependent variable, viz. $u\p_u$.

\section{The Klein-Gordon Equation in three dimensional Maximal Symmetric Spacetimes}
A Riemannian space which admits a Killing algebra of dimension $\frac12 n (n + 1)$ is called a maximally symmetric
metric space, for example, the Euclidian space $E^3$ and the Minkowski spacetime $M^3$,
 defined by \begin{equation} ds^{2}= \epsilon dt^{2}+dx^2 + dy^2, \quad \epsilon=\pm1.
 \label{eq11}\end{equation}
The conformal algebra of this space  is 10-dimensional and consists of the following vectors.

3- gradient KVs: $$X^1= \partial_t \quad X^2=\partial_{x}\quad X^3=\partial_{y},$$

1- gradient HV:  $$X^4=x\partial_x+y\partial_y+t\partial_t,$$

3- rotations:
$$\displaystyle{X^5={y}\, \partial_{x}-{x}\, \partial_{y},\quad X^6=-\epsilon t\, \partial_{x}+{ {{x}}{ }}\, \partial_t,\quad X^7=-\epsilon t\, \partial_{y}+{{{y}}{ }}\, \partial_t,}$$

3- special CKVs with their respective conformal factors:

\begin{equation} \begin{array}{lcl}
X^8&= 2{{x}{y}}\, \partial_{x}-\,({{t}^{2}\epsilon +{{x}}^{2}-{{y}}^{2}})\, \partial_{y}+2{{y}t}\, \partial_t,\qquad &\psi^1=2y\\
X^9&=-(\,{{t}^{2}\epsilon -{{x}}^{2}+{{y}}^{2}})\, \partial_{x}+2{x}{y}\, \partial_{y}+2{x}t\, \partial_t,\qquad &\psi^2=2x\\
X^{10}&\displaystyle{= 2{x}t\, \partial_{x}+2{y}t\, \partial_{y}+\,{\frac {{t}^{2}\epsilon -{{x}}^{2}-{{y}}^{2}}{{\epsilon }}}}\, \partial_t,\qquad &\psi^3=2t.
\nonumber\label{ckv}\end{array}\end{equation}
In Table \ref{tab:table000}, we list real subalgebras of the conformal algebra, where the algebra $F_{r,k}$ denotes the $k$-th algebra of dimension $r$ and excludes linear combinations.
\begin{table}[!ht]\caption{\label{tab:table000}%
Real subalgebras of the conformal algebra}\begin{adjustbox}{max width=\linewidth}
\begin{tabular}{lclclc}
\hline
\textrm{\textbf{Name}}&
\textrm{\textbf{Generators}}&
\multicolumn{1}{c}{\textbf{Name}}&
\textrm{\textbf{Generators}}&
\textrm{\textbf{Name}}&
\textrm{\textbf{Generators}}\\
\hline\hline
$F_{2,1}$ &$ X^8,X^{10} $  &$F_{2,2}$ & $ X^8,X^9 $ &$F_{2,3}$ &$ X^8,X^4 $\\
$ F_{2,4}$ &$ X^8,X^6 $ &$F_{2,5}$ &$ X^{10},X^9 $ &$F_{2,6}$ &$ X^{10},X^4 $\\
$F_{2,7}$ &$ X^{10},X^5 $&$F_{2,8}$ &$ X^9,X^4 $ &$F_{2,9}$ &$ X^9,X^7 $ \\
$F_{2,10}$ &$ X^4,X^2 $&$F_{2,11}$ &$ X^4,X^1 $&$F_{2,12}$ &$ X^4,X^5 $\\
$F_{2,13}$ &$ X^4,X^7 $&$F_{2,14}$ &$ X^4,X^3 $&$F_{2,15}$ &$ X^4,X^6 $ \\
$F_{2,16}$ &$ X^2,X^1 $&$F_{2,17}$ &$ X^2,X^7 $&$F_{2,18}$ &$ X^2,X^3 $ \\
$F_{2,19}$ &$ X^1,X^5 $&$F_{2,20}$ &$ X^1,X^3 $ &$F_{2,21}$ &$ X^3,X^6 $ \\
$F_{3,1}$ &$ X^8,X^{10},X^9 $&$F_{3,2}$ &$ X^8,X^{10},X^4 $ &$F_{3,3}$ &$ X^8,X^{10},X^7 $ \\
$F_{3,4}$ &$ X^8,X^9,X^4 $ &$F_{3,5}$ &$ X^8,X^9,X^5 $ &$F_{3,6}$ &$ X^8,X^4,X^3 $ \\
$F_{3,7}$ &$ X^8,X^4,X^6 $ &$F_{3,8}$ &$ X^{10},X^9,X^4 $ &$F_{3,9}$ &$ X^{10},X^9,X^6 $ \\
$F_{3,10}$ &$ X^{10},X^4,X^1 $ &$F_{3,11}$ &$ X^{10},X^4,X^5 $ &$F_{3,12}$ &$ X^9,X^4,X^2 $ \\
$F_{3,13}$ &$ X^9,X^4,X^7 $ &$F_{3,14}$ &$ X^4,X^2,X^1 $ &$F_{3,15}$ &$ X^4,X^2,X^7 $ \\
$F_{3,16}$ &$ X^4,X^2,X^3 $ &$F_{3,17}$ &$ X^4,X^1,X^5 $ &$F_{3,18}$ &$ X^4,X^1,X^3 $ \\
$F_{3,19}$ &$ X^4,X^3,X^6 $ &$F_{3,20}$ &$ X^2,X^1,X^3 $ &$F_{3,21}$ &$ X^2,X^1,X^6 $ \\
$F_{3,22}$ &$ X^2,X^5,X^3 $ &$F_{3,23}$ &$ X^1,X^7,X^3 $ &$F_{3,24}$ &$ X^5,X^7,X^6 $ \\
$F_{4,1}$ &$ X^8,X^{10},X^9,X^4 $ &$F_{4,2}$ &$ X^8,X^{10},X^9,X^5 $ & $F_{4,3}$ &$ X^8,X^{10},X^9,X^7 $\\
$F_{4,4}$ &$ X^8,X^{10},X^9,X^6 $ &$F_{4,5}$ &$ X^8,X^{10},X^4,X^7 $ &$F_{4,6}$ &$ X^8,X^9,X^4,X^5 $ \\
$F_{4,7}$ &$ X^8,X^4,X^3,X^6 $ &$F_{4,8}$ &$ X^{10},X^9,X^4,X^6 $  &$F_{4,9}$ &$ X^{10},X^4,X^1,X^5 $ \\
$F_{4,10}$ &$ X^9,X^4,X^2,X^7 $  &$F_{4,11}$ &$ X^4,X^2,X^1,X^3 $ &$F_{4,12}$ &$ X^4,X^2,X^1,X^6 $ \\
$F_{4,13}$ &$ X^4,X^2,X^5,X^3 $  &$F_{4,14}$ &$ X^4,X^1,X^7,X^3 $ &$F_{4,15}$ &$ X^4,X^5,X^7,X^6 $\\
$F_{4,16}$ &$ X^2,X^1,X^5,X^3 $&$F_{4,17}$ &$ X^2,X^1,X^7,X^3 $ &$F_{4,18}$ &$ X^2,X^1,X^3,X^6 $ \\
$F_{5,1}$ &$ X^8,X^{10},X^9,X^4,X^5 $  &$F_{5,2}$ &$ X^8,X^{10},X^9,X^4,X^7 $&&\\
 $F_{5,3}$ &$ X^8,X^{10},X^9,X^4,X^6 $ &$F_{5,4}$ &$ X^4,X^2,X^1,X^5,X^3 $&&\\
 $F_{5,5}$ &$ X^4,X^2,X^1,X^7,X^3 $ &$F_{5,6}$ &$ X^4,X^2,X^1,X^3,X^6 $&& \\
 $F_{6,1}$ &$ X^8,X^{10},X^9,X^5,X^7,X^6 $  &$F_{6,2}$ &$ X^8,X^{10},X^4,X^1,X^7,X_{9} $  &&\\
 $F_{6,3}$ &$ X^8,X^9,X^4,X^2,X^5,X_{9} $ &$F_{6,4}$ &$ X^{10},X^9,X^4,X^2,X^1,X^6 $  &&\\
 $F_{6,5}$ &$ X^2,X^1,X^5,X^7,X^3,X^6 $  &$F_{7,1}$ &$ X^4,X^{5},X^6,X^7,X^8,X^9,X^{10} $  &&\\
 $F_{7,2}$ &$ X^1,X^2,X^3,X^4,X^5,X^6,X^7 $  &&&&\\ \hline\hline
\end{tabular}\end{adjustbox}
\end{table}

\begin{table}[!ht]%
\caption{\label{tab:table00}%
Lie brackets $\left[X^i,X^j\right]$ of the conformal algebra.}\begin{adjustbox}{max width=\textwidth}
\begin{tabular}{c|c|c|c|c|c|c|c|c|c|c}
\hline
\textrm{\textbf{}}&
\textrm{\textbf{$X^1$}}&
\multicolumn{1}{c}{\textrm{\textbf{$X^2$}}}&
\textrm{\textbf{$X^3$}}&
\textrm{\textbf{$X^4$}}&
\textrm{\textbf{$X^5$}}&
\textrm{\textbf{$X^6$}}&
\textrm{\textbf{$X^7$}}&
\textrm{\textbf{$X^8$}}&
\textrm{\textbf{$X^9$}}&
\textrm{\textbf{$X^{10}$}}\\\hline
$X^{1}$  & 0 &0 &0 &$X^1$ &0 &$\epsilon X^2$ &$\epsilon X^3$ &$-2X^7$ &$-2X^6$ &$2X^4$\\
$X^{2}$  & 0 &0 &0 &$X^2$ &$X^3$ &$- X^1$ &$0$ &$-2X^5$ &$2X^4$ &$\frac{2X^6}{\epsilon}$\\
$X^{3}$ & 0 &0 &0 &$X^3$ &$-X^2$ &0 &$-X^1$ &$2X^4$ &$2X^5$ &$\frac{2X^7}{\epsilon}$\\
$X^{4}$  & $-X^1$ &$-X^2$ &$-X^3$ &0 &0 &0 &0 &$X^8$ &$X^9$ &$X^{10}$\\
$X^{5}$  & 0 &$-X^3$ &$X^2$ &0 &0 &$-X^7$ &$X^6$ &$X^9$ &$-X^8$ &0\\
$X^{6}$  &$-\epsilon X^2$ &$X^1$ &0 &0 &$X^7$ &0 &$-\epsilon X^5$ &0 &$\epsilon X^{10}$ &$-X^9$\\
$X^{7}$  &$-\epsilon X^3$ &0 &$X^1$ &0 &$-X^6$ &$\epsilon X^5$ &0 &$\epsilon X^{10}$ &0 &$-X^8$\\
$X^{8}$  &$2X^7$ &$2X^5$ &$-2X^4$ &$-X^8$ &$-X^9$ &0 &$-\epsilon X^{10}$ &0 &0 &0\\
$X^{9}$  &$2X^6$ &$-2X^4$ &$-2X^5$ &$-X^9$ &$X^8$ &$-\epsilon X^{10}$ &0 &0 &0 &0\\
$X^{10}$  &$-2X^4$ &$-\frac{2X^6}{\epsilon}$ &-$\frac{2X^7}{\epsilon}$ &$-X^{10}$ &0 &$X^9$ &$X^8$ &0 &0 &0\\\hline\hline
\end{tabular}\end{adjustbox}
\end{table}

The Klein-Gordon equation follows from the Lagrangian
\bn L(x^i,u,u_{,i})=\ds{\frac12 V(t,x,y) u^2-\frac12u^2_{,x}-\frac12u^2_{,y}-\frac{1}{2\epsilon}u^2_{,t},}
\label{eq333}\en
and is explicitly expressed as
\bn \frac1\epsilon u_{,tt}+ u_{,xx} +u_{,yy}+V(x,y,t) u=0.\label{eq33}\en

In this study, we apply elements of the conformal algebra in the solution of Eq. (\ref{pa}) to determine the  form of the potentials $V (t, x, y)$. That is, we consider three cases below, namely we apply (a)   the  vectors $X^{1-10}$,  (b) selected linear combinations of the vectors $X^{1-10}$ and (c) real subalgebras of the conformal algebra.
For each CKV of the conformal algebra of the  maximal symmetric spacetimes, we must solve constraint condition (\ref{pa}) where $ n = 3$ and find the potentials $V(x,y,t)$ for which it is satisfied. Before we proceed with the symmetry analysis, we mention that the Klein-Gordon equation is a linear equation which implies that it will always admit the linear symmetry $X^{11}= u\p_u$ and the infinite dimensional abelian subalgebra of solutions $X^\infty = F(x,y,t) \p_u$, where
$F(x,y,t)$  is a solution of Eq. (\ref{eq33}). Due to the many results and for ease of reference, the results are presented in the form of tables.

\subsection{Case I - The vectors $X^{1-10}$.}

Taking each of the vectors of the conformal algebra $X^{1-10}$, we solved Eq. (\ref{pa}) and found the potentials $V(t, x, y)$ of  Table \ref{tab:table1}.
The columns of Table \ref{tab:table1} contain the potential functions, corresponding Lie point symmetries, Lie invariant functions, Noether point symmetries and lastly, the associated conservation laws $T_{1-10}$ appear in Table \ref{tab:table5}. Hence, $(T^t,T^x,T^y)$ is the conserved vector that satisfies
$$
D_tT^t+D_xT^x+D_yT^y=0
$$
on Eq. (\ref{eq33}).
\begin{table}[!ht]%
\caption{\label{tab:table1}%
Point symmetries and potentials of the Klein-Gordon equation (\ref{eq33}) from Case I.}\begin{adjustbox}{max width=\textwidth}
\begin{tabular}{lccrr}
\hline
\textrm{\textbf{Potential}}&
\textrm{\textbf{Lie Symm.}}&
\multicolumn{1}{c}{\textrm{\textbf{Invariants}}}&
\textrm{\textbf{Noether Symm.}}&
\textrm{\textbf{Con. Law}}\\
\hline\hline
$V(t,x,y)$ & $X^{11}$ &$x, y, t$ &No &{-}\\
$V(x,y)$ & $X^1$ &$x, y, u$ &Yes &$T_1$\\
$V(t,y)$ & $X^2$ &$y, t, u$ &Yes &$T_2$\\
$V(t,x)$ & $X^3$ &$x, t, u$ &Yes &$T_3$\\
$\frac{1}{t^2}V\(\frac{x}{t},\frac{y}{t}\)$ & $X^4$ &$u, \frac{y}{x}, \frac{t}{x}$ &Yes &$T_4$\\
$V\(t,x^2+y^2\)$ & $X^5$ &$t, u, x^2+y^2$ &Yes &$T_5$\\
$V\(\epsilon t^2+x^2,y\)$ & $X^6$ &$y, u, \frac{\(t^2\epsilon+x^2\)}{\epsilon}$ &Yes &$T_6$\\
$V\(x,\epsilon t^2+y^2\)$ & $X^7$ &$x, u, \frac{\(t^2\epsilon+y^2\)}{\epsilon}$ &Yes &$T_7$\\
$\frac{1}{t^2}V\(\frac{x}{t},\frac{\epsilon t^2+x^2+y^2}{t}\)$ & $X^8+\frac12\psi^1 u\p_u$ &$\frac{t}{x}, u\sqrt{x}, \frac{\(t^2\epsilon+x^2+y^2\)}{x}$ &Yes &$T_8$\\
$\frac{1}{t^2}V\(\frac{y}{t},\frac{\epsilon t^2+x^2+y^2}{t}\)$ & $X^9+\frac12\psi^2 u\p_u$ &$\frac{t}{y}, u\sqrt{y}, \frac{\(t^2\epsilon+x^2+y^2\)}{y}$ &Yes &$T_9$\\
$\frac{1}{x^2}V\(\frac{y}{x},\frac{\epsilon t^2+x^2+y^2}{\epsilon x}\)$ & $X^{10}+\frac12\psi^3 u\p_u$ &$\frac{y}{x}, u\sqrt{x},\frac{\(t^2\epsilon+x^2+y^2\)}{x\epsilon}$ &Yes &$T_{10}$\\\hline\hline
\end{tabular}\end{adjustbox}
\end{table}

\begin{center}
\begin{longtable}{|l|l|}
\caption[Conservation Laws corresponding to $X^{1-10}$]{Conservation Laws corresponding to $X^{1-10}$.} \label{tab:table5} \\

\hline \multicolumn{1}{|c|}{$\dsy{T_j}$} & \multicolumn{1}{c|}{${T^t_j,T^x_j,T^y_j}$} \\ \hline\hline
\endfirsthead

\multicolumn{2}{c}%
{{\bfseries \tablename\ \thetable{} -- continued from previous page}} \\
\hline \multicolumn{1}{|c|}{$\dsy{T_j}$} &
\multicolumn{1}{c|}{${T^t_j,T^x_j,T^y_j}$}  \\ \hline\hline
\endhead

\hline \multicolumn{2}{|r|}{{Continued on next page}} \\ \hline\hline
\endfoot

\hline \hline
\endlastfoot
   $T_1 $&$ T^t= \frac{1}{2 \epsilon }\(\epsilon  u ^2 V(x,y)+\epsilon  u  \(u_{ yy}+u_{ xx}\)+u_{ t}{}^2\), \quad T^x= \frac{1}{2} \(u_{ x} u_{ t}-u  u_{ tx}\), \quad T^y= \frac{1}{2} \(u_{ y} u_{ t}-u  u_{ ty}\)$ \\
  \hline
  $T_2 $&$ T^t=\frac{1}{2 \epsilon }\(u_{ x} u_{ t}-u  u_{ tx}\),\quad T^x= \frac{1}{2 \epsilon }\(\epsilon  u ^2 V(t,y)+\epsilon  u_{ x}{}^2+u  \(\epsilon  u_{ yy}+u_{ tt}\)\), \quad T^y=\frac{1}{2} \(u_{ y} u_{ x}-u  u_{ xy}\) $\\
    \hline
    $T_3 $&$ T^t=\frac{1}{2 \epsilon }\(u_{ y} u_{ t}-u  u_{ ty}\), \quad T^x = \frac{1}{2} \(u_{ y} u_{ x}-u  u_{ xy}\), \quad T^y=\frac{1}{2 \epsilon }\(\epsilon  u ^2 V(t,x)+\epsilon  u_{ y}{}^2+u  \(\epsilon  u_{ xx}+u_{ tt}\)\)$ \\
    \hline
   $T_4 $&$ T^t= \frac{1}{2 t \epsilon }\(\epsilon  u ^2 V\(\frac{x}{t},\frac{y}{t}\)+t u_{ t} \(y u_{ y}+x u_{ x}+t u_{ t}\)+t u  \(t \epsilon  u_{ yy}+t \epsilon  u_{ xx}-u_{ t}-y u_{ ty}-x u_{ tx}\)\),$\\
   ${}$&$ T^x= \frac{1}{2 t^2 \epsilon }\(x \epsilon  u ^2 V\(\frac{x}{t},\frac{y}{t}\)+t^2 \epsilon  u_{ x} \(y u_{ y}+x u_{ x}+t u_{ t}\)-t^2 u  \(-x \epsilon  u_{ yy}+\epsilon  u_{ x}+y \epsilon  u_{ xy}+t \epsilon  u_{ tx}-x u_{ tt}\)\), $\\
   ${}$&$T^y= \frac{1}{2 t^2 \epsilon }\(y \epsilon  u ^2 V\(\frac{x}{t},\frac{y}{t}\)+t^2 \epsilon  u_{ y} \(y u_{ y}+x u_{ x}+t u_{ t}\)-t^2 u  \(\epsilon  u_{ y}+x \epsilon  u_{ xy}-y \epsilon  u_{ xx}+t \epsilon  u_{ ty}-y u_{ tt}\)\)$\\
  \hline
  $T_5$&$ T^t= \frac{1}{2 \epsilon }\(-x u_{ y} u_{ t}+y u_{ x} u_{ t}+u  \(x u_{ ty}-y u_{ tx}\)\),$\\
   ${}$&$ T^x= \frac{1}{2 \epsilon }\(y \epsilon  u ^2 V\(t,x^2+y^2\)+\epsilon  u_{ x} \(-x u_{ y}+y u_{ x}\)+u  \(\epsilon  u_{ y}+y \epsilon  u_{ yy}+x \epsilon  u_{ xy}+y u_{ tt}\)\), $\\
   ${}$&$T^y= -\frac{1}{2 \epsilon } \(x \epsilon  u ^2 V\(t,x^2+y^2\)+\epsilon  u_{ y} \(x u_{ y}-y u_{ x}\)+u  \(\epsilon  u_{ x}+y \epsilon  u_{ xy}+x \epsilon  u_{ xx}+x u_{ tt}\)\)$\\
     \hline
  $T_6$&$ T^t= \frac{1}{2 \epsilon }\(x \epsilon  u ^2 V\(x^2+t^2 \epsilon ,y\)+u_{ t} \(-t \epsilon  u_{ x}+x u_{ t}\)+\epsilon  u  \(x u_{ yy}+u_{ x}+x u_{ xx}+t u_{ tx}\)\),$\\
   ${}$&$ T^x= \frac{1}{2} \(-t \epsilon  u ^2 V\(x^2+t^2 \epsilon ,y\)+u_{ x} \(-t \epsilon  u_{ x}+x u_{ t}\)-u  \(t \epsilon  u_{ yy}+u_{ t}+x u_{ tx}+t u_{ tt}\)\), $\\
   ${}$&$T^y=\frac{1}{2} \(u_{ y} \(-t \epsilon  u_{ x}+x u_{ t}\)+u  \(t \epsilon  u_{ xy}-x u_{ ty}\)\)$\\
  \hline
  $T_7$&$ T^t=\frac{1}{2 \epsilon }\(y \epsilon  u ^2 V\(x,y^2+t^2 \epsilon \)+u_{ t} \(-t \epsilon  u_{ y}+y u_{ t}\)+\epsilon  u  \(u_{ y}+y u_{ yy}+y u_{ xx}+t u_{ ty}\)\),$\\
   ${}$&$ T^x=\frac{1}{2} \(-t \epsilon  u_{ y} u_{ x}+y u_{ x} u_{ t}+u  \(t \epsilon  u_{ xy}-y u_{ tx}\)\),$ \\
    ${}$&$T^y=\frac{1}{2} \(-t \epsilon  u ^2 V\(x,y^2+t^2 \epsilon \)+u_{ y} \(-t \epsilon  u_{ y}+y u_{ t}\)-u  \(t \epsilon  u_{ xx}+u_{ t}+y u_{ ty}+t u_{ tt}\)\)$\\
   \hline
  $T_8$&$ T^t=\frac{1}{2 t \epsilon }(2 y \epsilon  u ^2 V(\frac{x}{t},\frac{x^2+y^2+t^2 \epsilon }{t})+t u_{ t} (-(x^2-y^2+t^2 \epsilon ) u_{ y}+2 y (x u_{ x}+t u_{ t}))$\\
  ${}$&$+t u  (2 t \epsilon  u_{ y}+2 t y \epsilon  u_{ yy}+2 t y \epsilon  u_{ xx}-2 y u_{ t}+x^2 u_{ ty}-y^2 u_{ ty}+t^2 \epsilon  u_{ ty}-2 x y u_{ tx})),$\\
  ${}$&$ T^x=\frac{1}{2 t^2 \epsilon }(2 x y \epsilon  u ^2 V(\frac{x}{t},\frac{x^2+y^2+t^2 \epsilon }{t})+t^2 \epsilon  u_{ x} (-(x^2-y^2+t^2 \epsilon ) u_{ y}+2 y (x u_{ x}$\\
  ${}$&$+t u_{ t}))+t^2 u  (2 x \epsilon  u_{ y}+2 x y \epsilon  u_{ yy}-2 y \epsilon  u_{ x}+x^2 \epsilon  u_{ xy}-y^2 \epsilon  u_{ xy}+t^2 \epsilon ^2 u_{ xy}-2 t y \epsilon  u_{ tx}+2 x y u_{ tt})), $\\
   ${}$&$T^y=--\frac{1}{2 t^2 \epsilon }(\epsilon  u ^2 (t^2+(x^2-y^2+t^2 \epsilon ) V(\frac{x}{t},\frac{x^2+y^2+t^2 \epsilon }{t}))+t^2 \epsilon  u_{ y} ((x^2-y^2+t^2 \epsilon ) u_{ y}$\\
   ${}$&$-2 y (x u_{ x}+t u_{ t}))+t^2 u  (2 y \epsilon  u_{ y}+2 x \epsilon  u_{ x}+2 x y \epsilon  u_{ xy}+x^2 \epsilon  u_{ xx}-y^2 \epsilon  u_{ xx}$\\
   ${}$&$+t^2 \epsilon ^2 u_{ xx}+2 t \epsilon  u_{ t}+2 t y \epsilon  u_{ ty}+x^2 u_{ tt}-y^2 u_{ tt}+t^2 \epsilon  u_{ tt}))$\\
     \hline
   $T_9$&$ T^t=\frac{1}{2 t \epsilon }(2 x \epsilon  u ^2 V(\frac{y}{t},\frac{x^2+y^2+t^2 \epsilon }{t})+t u_{ t} (2 x y u_{ y}+(x^2-y^2-t^2 \epsilon ) u_{ x}+2 t x u_{ t})$\\
   ${}$&$+t u  (2 t x \epsilon  u_{ yy}+2 t \epsilon  u_{ x}+2 t x \epsilon  u_{ xx}-2 x u_{ t}-2 x y u_{ ty}-x^2 u_{ tx}+y^2 u_{ tx}+t^2 \epsilon  u_{ tx})),$\\
  ${}$&$ T^x=-\frac{1}{2 t^2 \epsilon }(\epsilon  u ^2 (t^2+(-x^2+y^2+t^2 \epsilon ) V(\frac{y}{t},\frac{x^2+y^2+t^2 \epsilon }{t}))-t^2 \epsilon  u_{ x} (2 x y u_{ y}$\\
  ${}$&$+(x^2-y^2-t^2 \epsilon ) u_{ x}+2 t x u_{ t})+t^2 u  (2 y \epsilon  u_{ y}+\epsilon  (-x^2+y^2+t^2 \epsilon ) u_{ yy}+2 x \epsilon  u_{ x}+2 x y \epsilon  u_{ xy}$\\
  ${}$&$+2 t \epsilon  u_{ t}+2 t x \epsilon  u_{ tx}-x^2 u_{ tt}+y^2 u_{ tt}+t^2 \epsilon  u_{ tt})), $\\
   ${}$&$T^y=\frac{1}{2 t^2 \epsilon }(2 x y \epsilon  u ^2 V(\frac{y}{t},\frac{x^2+y^2+t^2 \epsilon }{t})+t^2 \epsilon  u_{ y} (2 x y u_{ y}+(x^2-y^2-t^2 \epsilon ) u_{ x}$\\
   ${}$&$+2 t x u_{ t})+t^2 u  (-2 x \epsilon  u_{ y}+2 y \epsilon  u_{ x}-x^2 \epsilon  u_{ xy}+y^2 \epsilon  u_{ xy}+t^2 \epsilon ^2 u_{ xy}+2 x y \epsilon  u_{ xx}-2 t x \epsilon  u_{ ty}+2 x y u_{ tt}))$\\
   \hline
$T_{10}$&$ T^t=-\frac{1}{2 x^2 \epsilon ^2}(\epsilon  u ^2 (x^2+(x^2+y^2-t^2 \epsilon ) V(\frac{y}{x},\frac{x^2+y^2+t^2 \epsilon }{x \epsilon }))$\\
${}$&$+x^2 u_{ t} (-2 t y \epsilon  u_{ y}-2 t x \epsilon  u_{ x}+(x^2+y^2-t^2 \epsilon ) u_{ t})+x^2 \epsilon  u  (2 y u_{ y}+(x^2+y^2-t^2 \epsilon ) u_{ yy}+2 x u_{ x}$\\
${}$&$+x^2 u_{ xx}+y^2 u_{ xx}-t^2 \epsilon  u_{ xx}+2 t u_{ t}+2 t y u_{ ty}+2 t x u_{ tx})),$\\
  ${}$&$ T^x=\frac{1}{2 x \epsilon }(2 t \epsilon  u ^2 V(\frac{y}{x},\frac{x^2+y^2+t^2 \epsilon }{x \epsilon })+x u_{ x} (2 t y \epsilon  u_{ y}+2 t x \epsilon  u_{ x}-$\\
  ${}$&$(x^2+y^2-t^2 \epsilon ) u_{ t})+x u  (2 t x \epsilon  u_{ yy}-2 t \epsilon  u_{ x}-2 t y \epsilon  u_{ xy}+2 x u_{ t}+x^2 u_{ tx}+y^2 u_{ tx}-t^2 \epsilon  u_{ tx}+2 t x u_{ tt})), $\\
   ${}$&$T^y=\frac{1}{2 x^2 \epsilon }(2 t y \epsilon  u ^2 V(\frac{y}{x},\frac{x^2+y^2+t^2 \epsilon }{x \epsilon })+x^2 u_{ y} (2 t y \epsilon  u_{ y}+2 t x \epsilon  u_{ x}$\\
   ${}$&$-(x^2+y^2-t^2 \epsilon ) u_{ t})+x^2 u  (-2 t \epsilon  u_{ y}-2 t x \epsilon  u_{ xy}+2 t y \epsilon  u_{ xx}+2 y u_{ t}+x^2 u_{ ty}+y^2 u_{ ty}-t^2 \epsilon  u_{ ty}+2 t y u_{ tt}))$
   \end{longtable}
\end{center}

\subsection{Case II - Linear combinations of $X^{1-10}$.}

In this case, for linear combinations, we take pairs of each of the vectors of the conformal algebra $X^{1-10}$. In turn these linear combinations are applied to Eq. (\ref{pa}) and the potentials $V(t, x, y)$ of  Table \ref{tab:table2} are determined. Note that not {all pairs} of linear combinations of the vector fields  provide us with  potential functions. We do not consider all other possible linear combinations because the resulting Lie and Noether symmetries are too many but they can be computed in the standard way.  In Table \ref{tab:table2} below, $a$ and $b$ are arbitrary non-zero constants.

\begin{table}[!ht]
\caption{\label{tab:table2}%
Point symmetries and potentials of the Klein-Gordon equation (\ref{eq33}) from Case II.}\begin{adjustbox}{max width=1.2\linewidth}
\begin{tabular}{lcc}
\hline
\textrm{\textbf{Potential}}&
\textrm{\textbf{Lie Symmetry}}&
\textrm{\textbf{Noether Symmetry}}\\
\hline\hline
$V\(x-\frac{b}{a}t,y\)$ & $aX^1+bX^2$  &Yes \\
$V\(t,y-\frac{b}{a}x\)$ & $aX^2+bX^3$ &Yes\\
$V\(x,y-\frac{b}{a}t\)$ & $aX^1+bX^3$  &Yes\\
$\frac{1}{\(bt+a\)^2}V\(\frac{x}{bt+a},\frac{y}{bt+a}\)$ & $aX^1+bX^4$ &Yes \\
$\frac{1}{t^2}V\(\frac{bx+a}{bt},\frac{y}{t}\)$ & $aX^2+bX^4$  &Yes \\
$\frac{1}{t^2}V\(\frac{x}{t},\frac{by+a}{bt}\)$ & $aX^3+bX^4$ &Yes \\
$V\(x^2+y^2,\frac{tb-a \arctan\(\frac{x}{y}\)}{b}\)$ & $aX^1+bX^5$  &Yes \\
$V\(-\frac{\epsilon b t^2+bx^2+2ax}{2b\epsilon},y\)$ & $aX^1+bX^6$  &Yes \\
$V\(x,-\frac{\epsilon b t^2+by^2+2ay}{b\epsilon}\)$ & $aX^1+bX^7$  &Yes \\
$\frac{1}{x^2}V\(\frac{y}{x},\frac{\epsilon b t^2+bx^2+by^2+\epsilon a}{bx\epsilon}\)$ & $aX^1+b(X^{10}+\psi^3)$  &Yes \\
$V\(t,-\frac{bx^2+by^2+2ay}{2b}\)$ & $aX^2+bX^5$  &Yes \\
$V\(\frac{\epsilon b t^2+bx^2-2at}{b},y\)$ & $aX^2+bX^6$  &Yes \\
$V\(\epsilon t^2+y^2,\frac{\sqrt{\epsilon} b x-a \arctan{\(\frac{t\sqrt{\epsilon}}{y}\)}}{b\sqrt{\epsilon}}\)$ & $aX^2+bX^7$  &Yes\\
$\frac{1}{t^2}V\(\frac{y}{t},\frac{\epsilon b t^2+bx^2+by^2+ a}{bt}\)$ & $aX^2+b(X^{9}+\psi^2)$  &Yes \\
$V\(t,-\frac{bx^2+by^2-2at}{b}\)$ & $aX^3+bX^5$  &Yes\\
$V\(\epsilon t^2+x^2,\frac{\sqrt{\epsilon} b y-a \arctan{\(\frac{t\sqrt{\epsilon}}{x}\)}}{b\sqrt{\epsilon}}\)$ & $aX^3+bX^6$  &Yes\\
$V\(x,\frac{\epsilon b t^2+by^2-2at}{b}\)$ & $aX^3+bX^7$ &Yes\\
$V\(y+\frac{a}{b}t,\frac{\epsilon t^2b^2-2at(by+at)+x^2b^2+t^2a^2}{b^2}\)$ & $aX^5+bX^6$ &Yes\\
$V\(y-\frac{b}{a}x,\frac{\epsilon t^2a^2+2bx(ay-bx)+x^2a^2+x^2b^2}{a^2\epsilon}\)$ & $aX^6+bX^7$  &Yes\\
$V\(x-\frac{a}{b}t,\frac{\epsilon t^2b^2+2at(bx-at)+y^2b^2+t^2a^2}{b^2}\)$ & $aX^5+bX^7$  &Yes\\\hline\hline
\end{tabular}\end{adjustbox}
\end{table}

\subsection{Case III - Real subalgebras of $X^{1-10}$.}

The real subalgebras contained within the conformal algebra $X^{1-10}$  which solve condition Eq. (\ref{pa}), are used  in order to determine all the potentials in which the Klein-Gordon equation admits Lie and Noether symmetries. The list of potential functions appear in Table  \ref{grid} - \ref{grid1} together with the corresponding point symmetries.   It is important to note that we display the smallest subalgebra that admits potentials in Tables \ref{grid}. Moreover, we have that  $a_r,b_r,c_r\ne0\; (r=1,2,3,4,5)$ in Table  \ref{grid1} which include subalgebras containing linear combinations. It is necessary to remark for Case III, that when the Klein-Gordon equation admits a special CKV as its Lie/Noether point symmetry, then the form of the Lie point symmetry  is expressed as a sum of the special CKV and its respective conformal factor, i.e. for instance we would have $X^8+\frac12  \psi^1u  \p_u$.

\begin{table}[!ht]
\caption{\label{grid}%
Case III: The Lie subgroups  and its admitted potentials for Eq. (\ref{eq33}).}\begin{adjustbox}{max width=1.3\linewidth}
\begin{tabular}{lc}
\hline
\textrm{\textbf{Potential Function}}&
\textrm{\textbf{Lie/Noether Algebra}}\\
\hline\hline
$\frac{1}{t^2}V\(\frac{x^2+y^2}{t^2}\)$ & $F_{2,12}$ \\
$V(x^2+y^2)$; $V\(\frac{1}{x^2+y^2}\)$  & $F_{2,19}$ \\
$\frac{1}{x^2}V\(\frac{\epsilon t^2+y^2}{x^2}\)$ & $F_{2,13}$ \\
$\frac{1}{\epsilon t^2+x^2}V\(\frac{y}{\sqrt{\epsilon t^2+x^2}}\)$ & $F_{2,15}$ \\
$V\(\epsilon t^2+y^2\)$; $V\(\frac{1}{\epsilon t^2+y^2}\)$ & $F_{2,17}$ \\
$V\(\epsilon t^2+x^2\)$; $V\(\frac{1}{\epsilon t^2+x^2}\)$ & $F_{2,21}$ \\
$\frac{1}{\epsilon t^2+x^2}V\(\frac{\epsilon t^2+x^2+y^2}{\sqrt{\epsilon t^2+x^2}}\)$ & $F_{2,4}$ \\
$V(x)$; $V\(\frac{1}{x^2}\)$ & $F_{3,23}$ \\
$V(y)$; $V\(\frac{1}{y^2}\)$ & $F_{3,21}$ \\
$V(t)$; $V\(\frac{1}{t^2}\)$ & $F_{3,22}$ \\
$V\(\epsilon t^2+x^2+y^2\)$; $V\(\frac{1}{\epsilon t^2+x^2+y^2}\)$; $V\(\frac{1}{(\epsilon t^2+x^2+y^2)^2}\)$  & $F_{3,24}$ \\
$\frac{1}{t^2}V\(\frac{x}{t}\)$ & $F_{3,6}$ \\
$\frac{1}{x^2}V\(\frac{y}{x}\)$ & $F_{3,10}$ \\
$\frac{1}{t^2}V\(\frac{y}{t}\)$ & $F_{3,12}$ \\
$\frac{1}{x^2}V\(\frac{\epsilon t^2+x^2+y^2}{x}\)$ & $F_{3,3}$ \\
$\frac{1}{t^2}V\(\frac{\epsilon t^2+x^2+y^2}{t}\)$ & $F_{3,5}$ \\\hline\hline
\end{tabular}\end{adjustbox}
\end{table}
\begin{table}[htbp]
\caption{\label{grid1}%
Case III: The Lie algebra is spanned by linear combinations of the CKVs.}
\begin{tabular}{lc}
\hline
\textrm{\textbf{Potential Function}}&
\textrm{\textbf{Lie/Noether Algebra}}\\
\hline\hline
$V\((\epsilon a_1t-x)a_3+y\)$ & $\{\frac1\e X^1+a_1X^2,X^2+a_3X^{3}\}$ \\
$\frac{1}{\(\e a_1t-x\)^2}V\(\frac{y}{\e a_1t-x}\)$ & $\{\frac1\e X^1+a_1X^2,X^4\}$ \\
$V\(\frac{(\e^2t^2+\e y^2)a_1^2-2\e a_1tx+x^2+y^2}{\e a_1^2+1}\)$ & $\{\frac1\e X^1+a_1X^2,X^5-a_1X^{7}\}$ \\
$V\(-\epsilon a_1t+x\)$ & $\{\frac1\e X^1+a_1X^2+a_5X^3,X^{3}\}$ \\
$V\(-\epsilon a_2t-xa_3+y\)$ & $\{\frac1\e X^1+a_2X^3,X^2+a_3X^{3}\}$ \\
$\frac{1}{x^2}V\(\frac{-\e t a_2+y}{x}\)$ & $\{\frac1\e X^1+a_2X^3,X^4\}$  \\
$V\((\e^2t^2+\e x^2)a_2^2-2\e a_2ty+x^2+y^2\)$ & $\{\frac1\e X^1+a_2X^3,X^5+a_2X^6\}$ \\
$V\(-\epsilon a_2t+y\)$ & $\{\frac1\e X^1+a_4X^2+a_2X^3,X^2\}$ \\
$\frac{1}{t^2}V\(\frac{-a_3x+y}{t}\)$ & $\{X^2+a_3X^3,X^4\}$ \\
$V\(t(a_3a_4-a_5)\e-a_3x+y\)$ & $\{X^2+a_3X^3,X^1+a_4X^2+a_5X^3\}$ \\
$V\(\frac{(\e t^2+ y^2)b_3^2+2b_3xy+x^2+t^2\e}{b_3^2}\)$ & $\{X^2-\frac{1}{b_3}X^3,\frac1\e(X^6+b_3X^7)\}$ \\
$\frac{1}{\(\e a_4t-x\)^2}V\(\frac{-\e a_5t+y}{\e a_4t-x}\)$ & $\{\frac1\e X^1+a_4X^2+a_5X^3,X^{4}\}$ \\
$V\(\frac{\e (b_1^2t^2-2b_1ty+x^2+y^2)+x^2b_1^2}{\e }\)$ & $\{\frac1\e X^1+\frac{b_1}{\e}X^3,X^5+\frac{b_1}{\e}X^{6}\}$ \\
$V\(\frac{\e (b_2^2t^2+2b_2tx+x^2+y^2)+y^2b_2^2}{\e +b_2^2}\)$ & $\{\frac1\e X^1-\frac{b_2}{\e}X^2,X^5+\frac{b_2}{\e}X^{7}\}$ \\
$\frac{1}{\(\e t-b_2x\)^2}V\(\frac{(\e t^2+ y^2)b_2^2+2b_2\e xt-\e^2t^2}{b_2^2(-\e t+b_2x)^2}\)$ & $\{X^4,X^5+\frac{b_2}{\e}X^7\}$ \\
$\frac{1}{t^2}V\(\frac{yc_2-x}{c_2t}\)$ & $\{X^4,X^8+\frac{c_2}{2}X^{9}\}$ \\
$\frac{1}{(-c_5\e t+c_4x)^2}V\(\frac{\e t-c_4y}{c_4(-\e c_5 t+c_4x)}\)$ & $\{X^4,X^{8}+\frac{c_4}{2}X^{10}+\frac{c_5}{2}X^{9}\}$ \\
$\frac{1}{(c_3\e t-x)^2}V\(\frac{\epsilon t^2+x^2+y^2}{c_3\e t-x}\)$ & $\{X^8,X^{10}+{c_3}X^{9}\}$ \\
$\frac{1}{x^2}V\(\frac{-\e t+c_1y}{c_1x}\)$ & $\{X^4,\frac1\e X^1+\frac{1}{c_1}X^3,X^8+\frac{c_1}{2}X^{10}\}$ \\
$\frac{1}{(c_3\e t^2-x)^2}V\(\frac{y}{c_3\e t-x}\)$ & $\{X^4,\frac1\e X^1+c_3X^2,X^{10}+\frac{c_3}{2}X^{9}\}$ \\
$\frac{c}{(\epsilon a_1t-x)a_3+y}$ & $\{\frac1\e X^1+a_1X^2,X^4,X^2+a_3X^{3}\}$ \\
$V\(-a_3x+y\)$ & $\{X^1,X^2+a_3X^3,\frac1\e(X^6+a_3X^7)\}$ \\
$\frac{c}{(\epsilon a_2t+a_3x-y)^2}$ & $\{\frac1\e X^1+a_2X^3,X^4,X^2+a_3X^{3}\}$ \\
$V\(\frac{\e t+b_1y}{b_1}\)$ & $\{X^1-\frac{1}{b_1}X^3,X^2,X^5+\frac{b_1}{\e}X^6\}$ \\
$V\(\frac{c}{(\e^2t^2+\e y^2)a_1^2-2\e a_1tx+x^2+y^2}\)$ & $\{X^1+{a_1}X^2,X^4,X^5-{a_1}X^6\}$ \\
$V\(\frac{c}{(\e^2t^2+\e x^2)a_2^2-2\e a_2ty+x^2+y^2}\)$ & $\{X^1+{a_2}X^3,X^4,X^5+{a_2}X^6\}$\\
$V(\e^2t^2(a_4^2+a_5^2)+\e (y^2a_4^2+a_5^2x^2)-$ & $\{\frac1\e X^1+a_4X^2+a_5X^3,X^5+{a_5}X^{6}-a_4X^7\}$\\
$\phantom{ww}2\e (a_4a_5xy+ a_4tx+ a_5ty)+x^2+y^2)^{\frac12}$ & $\phantom{ww}$ \\\hline\hline
\end{tabular}
\end{table}

\section{Invariant Solutions}
The above tables are useful because they provide the appropriate Lie point symmetries which can be used for the reduction of the Klein-Gordon equation and subsequently the determination of corresponding invariant solutions. In this section, we apply the Lie symmetries in order to reduce Eq. (\ref{eq33}). We study the two cases: $V(t,x,y)=V(\epsilon t^2+y^2)$ and $V(t,x,y)=V(-a_3x+y)$.

${\bf a. V(t,x,y)=V(\epsilon t^2+y^2)}.$ Based on Table \ref{grid}, the subgroup labeled $F_{2,17}$ admits this particular potential.
Thus, for the purpose of Lie reduction we may utilise the symmetries
$$Y_1=X^2+\kappa_1 X^{11}\quad \textrm{and}\quad Y_2=X^7+\kappa_2X^{11},$$ with Lie Bracket $\[Y_1,Y_2\]=0.$  Reduction with respect to the Lie invariants of the symmetry vector $Y_1$ gives
\bn\label{r1}u(t,x,y)=\textrm{exp}(\kappa_1 x)\zeta(t,y), \en where $\zeta(t,y)$ satisfies the equation
\bn\label{r2} \zeta_{,tt}+\epsilon \(\zeta_{,yy}+ \(\kappa_1^2+V(\epsilon t^2+y^2)\)\zeta \)=0.\en To this equation we apply $Y_2$ and obtain the second-order ordinary differential equation
\bn\label{r3}\(\kappa_2^2+2\epsilon\sigma\(\kappa_1^2+V(2\sigma)\) \)\phi+2\epsilon\sigma\(2\phi_{,\sigma}+2\sigma\phi_{,\sigma\sigma}\)=0, \en where $\sigma=\frac12\(\epsilon t^2+y^2\)$, $\phi(\sigma)$ and
\bn\label{r4}u(t,x,y)=\frac{1}{\sqrt{\epsilon}}\textrm{exp}\(\kappa_1 x\pm\kappa_2\arctan\(t\sqrt{\frac{\epsilon}{y^2}}\)\)\phi\(\frac12\(\epsilon t^2+y^2\)\). \en We continue with the determination of the invariant solutions for a second potential function.

${\bf b. V(t,x,y)=V(-a_3x+y)}.$ From Table \ref{grid1}, this potential function is admitted by the subgroup $\{X^1,X^2+a_3X^3,\frac1\e(X^6+a_3X^7),X^{11},X^\infty\}$.
Hence, for  reduction we may utilise the symmetries
$$Z_1=X^1+\kappa_3 X^{11}\quad \textrm{and}\quad Z_2=X^2+a_3X^3+\kappa_4X^{11},$$ with Lie Bracket $\[Z_1,Z_2\]=0.$  Reduction with respect to the Lie invariants of the symmetry vector $Z_1$ provides the solution
\bn\label{r5}u(t,x,y)=\textrm{exp}(\kappa_3 t)\beta(x,y), \en where $\beta(t,y)$ satisfies the equation
\bn\label{r6} \epsilon\(\beta_{,xx}+\beta_{,yy}\)+\epsilon V(-a_3x+y)\beta+\kappa_3^2=0.\en To Eq. (\ref{r6}) we apply $Z_2$ and obtain the second-order ordinary differential equation
\bn\label{r7}\(\kappa_3^2+\epsilon\(\kappa_4^2+V(\alpha)\) \)\rho+\epsilon\(-2a_3\kappa_4\rho_{,\alpha}+\(1+a_3^2\)\rho_{,\alpha\alpha}\)=0, \en where $\alpha=-a_3x+y$, $\rho(\alpha)$ and the solution of the Klein-Gordon equation (\ref{eq33}) is
\bn\label{r8}u(t,x,y)=\textrm{exp}\(\kappa_3 t+\kappa_4x\)\rho(-a_3x+y). \en

\section{Conclusion}
We have determined the functional form of potential functions for which the resulting Klein-Gordon equation in Euclidean and Minkowski three dimensional space admits Lie and Noether point symmetries. The application of the conformal Killing algebra  produces the classification of Lie and Noether point symmetries and potential functions. It is easily seen from Tables \ref{tab:table1} and \ref{tab:table2}, that the  generators of the Lie and Noether point symmetries are the CKVs and their linear combinations; while the real subalgebras generate interesting forms of the potential function in Table  \ref{grid}. Moreover, a consideration of subalgebras consisting of linear combinations of the vector fields produces a further list of potential functions in Table  \ref{grid1}.

Naturally each of the Noether symmetries listed here  can be used to determine a corresponding conservation law as displayed in Table \ref{tab:table5}. The usefulness of the tabular results can be seen in the   reduction of the (1+2)-dimensional Klein-Gordon equation. We  applied the zero-order invariants of the Lie symmetries and  reduced the Klein-Gordon equation to a linear ordinary differential equation. In particular, concerning $V(t,x,y)=V(\epsilon t^2+y^2)$ and $V(t,x,y)=V(-a_3x+y)$, we found the closed forms of the group invariant solutions. In the analysis of $V(t,x,y)=0$,  the Klein-Gordon equation becomes the wave equation which has a Lie symmetry algebra that is identically the conformal algebra $X^{1-10}$, and for a constant potential, i.e. $V(t,x,y)=V_0$, the Lie symmetry algebra contains $X^{1-3,5-7}$  plus the  the linear and infinite symmetry in both cases.

The results of this analysis can be used in various ways, such as to construct conservation laws for the equation of motions of a particle in the classical or the semi-classical approach. Last but not least, the results of this analysis holds for all the Yamabe equations (conformal Laplace equations) in which the underlying geometry, the metric which defines the Laplace operator, is conformally flat.

\textbf{Acknowledgments:}
SJ would like to acknowledge the financial support from the National Research Foundation of South Africa with grant number 99279. AP acknowledges the financial support of FONDECYT grant no. 3160121.

\nocite{*}

\section*{References}

\bibliography{jgp}

\end{document}